\documentstyle[12pt]{article}

\begin{document}

\begin{center}
{\LARGE\bf Comment on\\
``Neutron-Proton Spin-Correlation Parameter $A_{zz}$ at 68 MeV''
}
\\
\vspace*{.5cm}
{\large\sc R. Machleidt}
\\
{\it Department of Physics, University of Idaho,
 Moscow, ID 83843, USA}
\\
and
\\
{\large\sc I. Slaus}
\\
{\it Rudjer Boskovic Institute, POB 1016, 41001 Zagreb, Croatia}
\\
\today
\end{center}
\vspace*{.5cm}

\begin{abstract}
We present two arguments indicating
 that the large value for the $\epsilon_1$ mixing parameter
at 50 MeV, which the Basel group extracted from their recent
$A_{zz}$ measurement,
may be incorrect.
First, there are nucleon-nucleon (NN) potentials which predict
the $\epsilon_1$ at 50 MeV substantially below the Basel value
and reproduce the
Basel $A_{zz}$ data accurately.
Second, the large value for $\epsilon_1$
at 50 MeV proposed by the Basel group can only be explained by a
model for the NN interaction which is very unrealistic
(no $\rho$-meson and essentially a point-like
$\pi NN$ vertex) and overpredicts the $\epsilon_1$
in the energy range where it is well determined (150--500 MeV)
by a factor of two.
\end{abstract}

PACS numbers: 21.30.+y, 13.75.Cs, 13.88.+e, 25.10.+s
\vspace*{1cm}


In a recent Letter~\cite{Ham91}, a group at the University of Basel
reported a measurement of the spin correlation parameter $A_{zz}$
in neutron-proton scattering at 67.5 MeV. They also present results
from a phase shift analysis in which these new data play a crucial
role, particularly, for the $^3S_1-^3D_1$ mixing parameter $\epsilon_1$
for which they obtain the (rather large) value of $2.9^0\pm 0.3^0$
at 50 MeV.
It is the purpose of this Comment to point out that this large
value for $\epsilon_1$ at 50 MeV may be incorrect.
We present essentially
two arguments.

{\it First, there are nucleon-nucleon (NN) potentials which predict
the $\epsilon_1$ at 50 MeV as low as $\approx 1.5^0$ and reproduce the
Basel $A_{zz}$ data accurately.}
In Fig.~1 we compare the Basel $A_{zz}$ data
with the predictions by
the Reid~\cite{Rei68} ($2.36^0$), Nijmegen~\cite{NRS78} ($2.27^0$),
Paris~\cite{Lac80} ($1.89^0$), and Bonn~A~\cite{Mac89} ($1.55^0$)
potential (with the predictions for $\epsilon_1$ at 50 MeV in
parenthesis), which fit the data with a $\chi^2$/datum
of 116.7, 47.7, 1.55, and 1.72, respectively.
In contrast to the Basel claim,
{\it the models with a small $\epsilon_1$, namely Paris and
Bonn~A, fit the $A_{zz}$ data best.}
$A_{zz}$ is also sensitive to the $^1P_1$ phase shift
which at 50 MeV is predicted
to be $-10.95^0$ and $-11.05^0$ by Paris and Bonn~A, respectively;
the Basel group uses $-9.4^0$.
The $^1P_1$ phase shift is essentially determined by $np$
differential cross section data at backward
angles. The most recent
and very accurate $np$ $d\sigma/d\Omega$ backward angle data at 50 MeV
taken by the Karlsruhe group~\cite{Fin90} are reproduced with a
$\chi^2$/datum of 0.2 and 0.8 by Paris and Bonn~A,
respectively
(Nijmegen: 1.2; Reid: 8.7).
Finally, $A_{yy}$ must be considered, since for $A_{yy}$ the correlation
between $\epsilon_1$
and $^1P_1$ is of opposite sign as compared to $A_{zz}$. Paris and Bonn~A
fit the world data on $A_{yy}$ at 50 MeV
quite satisfatorily with a $\chi^2$/datum of
1.37 and 1.27, respectively.
The world data on all $np$ observables
 at 50 MeV are reproduced with a $\chi^2$/datum
of 1.6 by Paris and 1.4 by Bonn A.

{\it Second}, the large value for $\epsilon_1$
at 50 MeV proposed by the Basel group can only be explained by a
model for the NN interaction which is very unrealistic
(no $\rho$-meson and essentially a point-like
$\pi NN$ vertex) and overpredicts the $\epsilon_1$
in the energy range where it is well determined (150--500 MeV)
by a factor of two (see Fig.~2, dotted line `No $\rho$').
In contrast to the situation below 100 MeV, recent phase shift
analyses~\cite{Bug90,Arn93} agree very well in their
determinations of the $\epsilon_1$ in the energy range
150--500 MeV (cf.\ Fig.~2).
Realistic NN potentials which reproduce this well-determined part
of $\epsilon_1$ correctly (solid lines in Fig.~2) predict $\epsilon_1$
at 50 MeV substantially below the Basel point.
Even the Reid potential, which overpredicts
$\epsilon_1$ at intermediate energies (dashed line in Fig.~2),
is still below the Basel point.
Only a model that contains no $\rho$-meson exchange and applies
essentially no $\pi NN$ form factor (cutoff mass 10 GeV) can reproduce
the Basel result for $\epsilon_1$.
Note, however, that the predictions by such a model
for the deuteron and for most
phase parameters (besides $\epsilon_1$, particularly,
$^1P_1$ and $^3P_J$) are totally wrong; thus, it is unrealistic.

A new phase-shift analysis~\cite{KSS92}
which includes the Basel data reports the value $2.2^0\pm 0.5^0$
for $\epsilon_1$ at 50 MeV,
in agreement with the implications of this Comment.

This work was supported in part by the US National Science
Foundation (PHY-9211607).

\pagebreak

\begin{center}
\large\bf
Figure Captions
\end{center}

{\bf Figure~1.} The neutron-proton $A_{zz}$ data at 67.5 MeV
taken by the Basel group~\cite{Ham91} (solid dots) are compared
to predictions by potential models, which predict $\epsilon_1$
at 50 MeV as given in parenthesis.
[Bonn~A: solid line; Paris: long dashes.]

{\bf Figure~2.} Phase shift analysis results (solid dots
Ref.~\cite{Bug90}, solid triangles Ref.~\cite{Arn93})
and potential
model predictions
for the $\epsilon_1$ mixing parameter as discussed in the text.
The crossed diamond denotes the value claimed by the Basel group.

\newpage

\vspace*{3cm}

The figures, which are crucial for a proper understanding of this Comment,
are available upon request from
\begin{center}
{\sc machleid@tamaluit.phys.uidaho.edu}
\end{center}
Please, include your FAX-number with your request.

\end{document}